%% file: main.tex
\documentclass[reprint, amsmath, amssymb, aps, prl]{revtex4-2}

\usepackage{graphicx}
\usepackage{tikz} % Graphics 
\tikzset{
  every node/.append style={
    execute at begin node={\everymath{\displaystyle}}
  }
}
\usepackage{dcolumn}
\usepackage{bm}
\usepackage{ulem}
\usepackage{xcolor}
\usepackage{slashed}
\usepackage{hyperref}
\usepackage{afterpage} % For using [!t] in the figure

\def\Tr{\text{Tr}}

\begin{document}

\title{Persistent Spontaneous Time-Reversal Breaking}

\author{Michael Smolkin}
\email{michael.smolkin@mail.huji.ac.il}

\author{Lev Yung}
\email{lev.yung@mail.huji.ac.il}

\affiliation{Racah Institute of Physics, The Hebrew University, Jerusalem 9190401, Israel}

\date{\today}

\begin{abstract}
We provide an analytic realization of spontaneous parity and time-reversal breaking without thermal restoration in a local, unitary, ultraviolet-complete quantum field theory in $2+1$ dimensions with finitely many fields.
The theory is defined by a relevant Yukawa deformation of a product of the critical biconical vector model and free massless Dirac fermions. This deformation induces a renormalization group flow whose infrared endpoint is a decoupled product of the critical vector model and the conformal Gross–Neveu–Yukawa model. At zero temperature, the vacuum preserves both symmetries, whereas upon heating, the theory undergoes an inverse thermal transition at a finite critical temperature. Above this temperature, a fermion mass, odd under parity and time reversal, is dynamically generated, and the broken phase persists to arbitrarily high temperatures.
\end{abstract}

\maketitle

%%%%%% Main Text %%%%%%
\section{Introduction}

Parity and time-reversal invariance play a fundamental role in our understanding of the laws of physics~\cite{Weinberg1995}. They provide important means of distinguishing and characterizing different phases of a theory. Unlike continuous symmetries, they are not generated by continuously varying transformations. Nevertheless, their presence or absence imposes powerful constraints on the dynamics, the spectrum, and the structure of correlation functions. The violation of parity and time-reversal symmetry in the weak interactions famously demonstrates that spacetime reflections need not be exact symmetries of nature~\cite{Lee:1956qn,Wu:1957my,Christenson:1964fg}. While they remain exact symmetries of the microscopic Lagrangian in many quantum field theories, they may nevertheless be broken explicitly by the dynamics or spontaneously by the state, either at zero or finite temperature, with potentially important consequences for the infrared behavior.

The realization of parity and time reversal is especially interesting in strongly interacting quantum field theories, where their fate may depend nontrivially on the dimensionality and structure of the interactions. Moreover, quantum effects can lead to subtleties in maintaining these symmetries even when they are manifest at the classical level. Understanding when and how discrete spacetime symmetries are preserved or broken therefore provides a useful window into the interplay between symmetry, dynamics, and infrared physics.

A particularly striking manifestation of discrete symmetry breaking in condensed matter is provided by Hall phenomena. The Hall response is odd under time reversal and therefore provides a direct experimental probe of time-reversal symmetry breaking. In the anomalous Hall effect, broken time-reversal symmetry gives rise to a transverse electrical response even in the absence of an external magnetic field, while the quantum anomalous Hall effect realizes a quantized version of this response~\cite{Haldane:1988,Nagaosa:2009ycg,Chang:2023}. These phenomena establish a direct connection between symmetry breaking, topological properties of quantum matter, and experimentally measurable transport coefficients. Since thermal fluctuations and symmetry breaking are intimately intertwined, it is natural to ask whether parity or time reversal, once broken at low temperatures, must be restored at sufficiently high temperatures. Persistent spontaneous symmetry breaking (PSSB) refers to the alternative, in which a symmetry is spontaneously broken in the high-temperature limit, either because the low-temperature order survives or because symmetry breaking emerges upon heating.

The possibility of symmetry breaking at high temperatures has a long history, dating back to the mechanism proposed by Weinberg~\cite{weinberg1974gauge} and subsequent studies of symmetry non-restoration~\cite{Fujimoto:1984hr,Salomonson:1984px,Klimenko:1988ng,Grabowski:1990qc,Roos:1995vm,Orloff:1996yn,Gavela:1998ux,Bimonte:1998he,Jansen:1998rj,Bimonte:1999tw,Pinto:1999pg}. These models, however, are not UV complete, preventing a controlled study of their high-temperature behavior at arbitrarily large temperatures. More recently, a variety of UV-complete relativistic quantum field theories exhibiting persistent symmetry breaking have been constructed~\cite{Chai:2020zgq,Chai:2020onq,Liendo:2022bmv,Chaudhuri:2020xxb,Bajc:2020gpa,Chai:2021djc,Chaudhuri:2021dsq,Chai:2021tpt,Hawashin:2024dpp,Komargodski:2024zmt,SmolkinYung:2025parity,Bajc:2026ppk,Chaudhuri:2026ges}. Developments in lattice systems have further clarified the mechanism underlying such high-temperature order in terms of entropic ordering: ordering in one sector can increase the entropy available to another, rendering the ordered phase thermodynamically favored even at arbitrarily high temperatures~\cite{Huang:2025gvi,Han:2025eiw,Andriolo:2026udg}. By contrast, existing holographic constructions with thermal order have so far exhibited only metastable ordered phases, with the true thermal equilibrium state remaining disordered~\cite{Buchel:2020thm,Buchel:2020jfs,Buchel:2021ead,Buchel:2022zxl,Buchel:2023zpe,Buchel:2025cve,Buchel:2025jup}.

In this paper, we provide analytic evidence for persistent parity and time-reversal breaking in a UV-complete, local, and unitary quantum field theory in $2+1$ dimensions with finitely many fields. The model was introduced recently in~\cite{SmolkinYung:2025parity}, where its phase structure was investigated using numerical functional renormalization group (FRG) methods directly in $2+1$ dimensions, supplemented by an $\epsilon$-expansion analysis. Those results provided numerical evidence that the symmetries remain spontaneously broken at arbitrarily high temperatures. Here we revisit the same model from a straightforward analytic perspective and demonstrate that the mechanism underlying persistent parity and time-reversal breaking can be understood directly from the structure of its effective potential.

In particular, we derive the finite-temperature effective potential and the critical temperature of the inverse thermal transition in a controlled limit, and show that the symmetry-breaking minimum persists to arbitrarily high temperatures. This establishes the persistence of the broken phase without relying on a numerical truncation of the FRG flow, which does not provide fully controlled error estimates. Our analysis also clarifies the role played by the interactions that stabilize the symmetry-broken phase and provides a fully analytic counterpart to the numerical evidence of~\cite{SmolkinYung:2025parity}. Together, these results strengthen the case that persistent parity and time-reversal breaking is a genuine feature of the theory rather than an artifact of a particular approximation scheme.

By construction, the model explicitly breaks conformal symmetry. It is defined by an RG flow between two fixed points. In the UV, it approaches a conformal field theory (CFT) consisting of free massless Dirac fermions decoupled from the critical biconical vector model~\cite{Fisher_mcl:1974,Vicari:2003}. The latter has been shown to exhibit PSSB using the $\epsilon$-expansion~\cite{Chai:2020zgq,Chai:2020onq}, truncated FRG methods~\cite{Hawashin:2024dpp}, and large-$N$ techniques~\cite{Komargodski:2024zmt,Smolkin:2026wij}. In the IR, the model flows to a product of two known CFTs: the Gross--Neveu--Yukawa (GNY) model and the critical vector model~\cite{Moshe:2003xn}. The flow is triggered by a relevant Yukawa-type deformation.

Since the Yukawa interaction introduces a dimensionful scale, the mere existence of a parity-odd fermion mass at finite temperature does not establish that thermal effects are responsible for symmetry breaking. The theory could already be in a gapped, symmetry-broken phase at $T=0$, in which case the persistence of a fermion mass at finite temperature would simply reflect the persistence of the broken phase, with thermal fluctuations failing to restore the symmetry. Such persistence, while interesting, would constitute a weaker statement than the phenomenon we demonstrate here.

In our construction, both the UV and IR limits are scale invariant. The corresponding CFTs are manifestly invariant under parity and time reversal, and the theory has no symmetry-breaking order at zero temperature. This order therefore emerges at finite temperature rather than persisting from a pre-existing broken phase. We demonstrate that it survives to arbitrarily high temperatures, establishing persistent symmetry breaking in the stronger sense relevant here.

\section{The Biconical Model}

We begin by reviewing the critical biconical model, which was recently shown to exhibit $\mathbb Z_2$ symmetry breaking that persists to arbitrarily high temperatures using a variety of approaches, including the $\epsilon$-expansion \cite{Chai:2020zgq,Chai:2020onq}, FRG \cite{Hawashin:2024dpp}, and large-$N$ techniques \cite{Komargodski:2024zmt, Smolkin:2026wij}. Our presentation also provides a simple bridge between the original calculation of \cite{Komargodski:2024zmt} and the direct computation of \cite{Smolkin:2026wij}. The critical biconical model plays a central role in the RG flow studied in the next section.

Consider a $d$-dimensional Euclidean field theory containing an $N$-component real scalar field $\phi_i$ and a single real scalar field $\chi$, described by the following $O(N)\times\mathbb{Z}_2$-symmetric UV action 
\begin{equation} 
\label{eq:bicon_uv_action}
\begin{split}
S = \int d^d x \bigg\{&
\frac12(\partial\phi)^2
+\frac12(\partial\chi)^2
+\frac{\lambda_\phi}{8N}\,(\phi^2)^2
\\&\,+\frac{\lambda_{\phi\chi}}{4N}\,\phi^2\,\chi^2
+\frac{\lambda_\chi}{8N}\,\chi^4 + \frac{g}{6! N^2 }\chi^6\bigg\},
\end{split}
\end{equation}
where $3\leq d <4$, the quartic couplings $\lambda_\phi, \lambda_{\phi\chi}$, and $ \lambda_{\chi}$ have mass dimension $4-d$, and the sextic coupling $g$ has mass dimension $6-2d$. For our purposes, it is sufficient to restrict the analysis to $\lambda_\phi > 0$. In the case of a zero sextic coupling, boundedness of the UV potential requires $\lambda_\chi \geq 0$ and $\lambda_{\phi\chi} \geq -\sqrt{\lambda_\phi \lambda_\chi}$. The quadratic mass terms are tuned to zero.

The theory admits a systematic $1/N$ expansion formulated entirely in terms of the original fields~\cite{Smolkin:2026wij}. In the current study, however, it is convenient to make the large-$N$ saddle-point structure explicit via the Hubbard--Stratonovich transformation. We introduce an auxiliary field $\sigma$, whose functional integration contour runs along the imaginary axis, and change the UV Lagrangian by a term that does not affect the dynamics of the theory,
$$\mathcal{L} \rightarrow\mathcal{L} - \frac{1}{2 \lambda_\phi} \left(\sigma -\frac{\lambda_\phi}{2\sqrt{N}}\phi^2-\frac{\lambda_{\phi\chi}}{2\sqrt{N}}\chi^2\right)^2.$$ 
Indeed, performing the Gaussian integral over $\sigma$ recovers the original theory up to a field-independent constant. Expanding the square, we obtain
\begin{equation}\label{eq:bicon_model_hs}
\begin{split}
    S = \int d^d x \bigg\{\frac{1}{2}(\partial \phi)^2 + \frac{1}{2}(\partial \chi)^2 + \frac{1}{2\sqrt{N}}\sigma(\phi^2 + \alpha \chi^2)\\ - \frac{\sigma^2}{2\lambda_\phi} + \frac{\kappa}{8N}\chi^4 + \frac{g}{6!N^2}\chi^6 \bigg\},
\end{split}
\end{equation}
where $\alpha = \lambda_{\phi\chi}/\lambda_\phi$ and $\kappa = \lambda_\chi -\lambda_{\phi\chi}^2/\lambda_\phi$. The two-point function of the $\sigma$ field at leading order (LO) in $1/N$ is given by the geometric resummation of the bubble chain,
\begin{center}

\input{pics/bubble_chain}
\end{center}
where the dashed and solid lines denote the bare $\sigma$ and $\phi$ propagators, respectively, and $\Pi(p)$ denotes the massless bosonic bubble, also known as the fish diagram,
\begin{equation} \label{eq:bubble}
    \Pi(p) = \frac{1}{2}\int_k \frac{1}{k^2(k+p)^2}, \qquad \int_k \equiv \int \frac{d^d k}{(2\pi)^d}.
\end{equation}
The integral evaluates to 
\begin{equation}
    \Pi(p) = a^{-1}|p|^{d-4}, \quad a =\frac{2(4\pi)^{d/2}\Gamma(d-2)}
{\Gamma\left(2-\frac d2\right)\Gamma\left(\frac d2-1\right)^2}.
\end{equation}
In the IR limit $|p|\ll \lambda_\phi^{\frac{1}{4-d}}$, the $\sigma$ propagator simplifies~to 
\begin{equation}
    \langle \sigma(p)\sigma(q) \rangle_{\tiny\text{IR}} = -(2\pi)^d \delta^{(d)}(p+q) \cdot a|p|^{4-d} + O(N^{-1}).
\end{equation}
By contrast, the two-point function of $\chi$ takes the free-field form at LO. The infrared scaling dimensions of $\sigma$ and $\chi$ are therefore
\begin{equation}
    \Delta_\sigma = 2 + O(N^{-1}), \qquad \Delta_\chi = \frac{d-2}{2} + O(N^{-1}).
\end{equation}
Large-$N$ factorization suppresses interactions among these operators, so at LO the scaling dimensions of composite operators constructed from them are additive. Consequently, the $\sigma^2$ term in \eqref{eq:bicon_model_hs}, as well as the $\chi^6$ term for $d>3$, are irrelevant, and the corresponding couplings vanish at the IR fixed point. In contrast, the $\chi^4$ term is relevant and drives the theory away from criticality. We therefore fine-tune the quartic couplings such that $\kappa=0$. With this tuning, the theory~\eqref{eq:bicon_uv_action} flows in the IR to the critical biconical model, whose action reads
\begin{equation}\label{eq:critical_bicon_model}
\begin{split}
    S = \int d^d x \bigg\{\frac{1}{2}(\partial \phi)^2 + \frac{1}{2}(\partial \chi)^2 + \frac{1}{2\sqrt{N}}\sigma(\phi^2 + \alpha \chi^2)\\+ \delta_{d,3}\frac{g}{6!N^2}\chi^6 \bigg\},
\end{split}
\end{equation}
where $\delta_{d,3}=1$ for $d=3$ and vanishes otherwise. This representation of the critical biconical model was studied in \cite{Komargodski:2024zmt}. At LO, this action defines a continuous family of CFTs parameterized by real $\alpha$ and, in $d=3$, by positive~$g$. The next-to-leading order (NLO) analysis~\cite{Komargodski:2024zmt, Smolkin:2026wij} reveals that the conformal manifold reduces to a set of isolated fixed points. The biconical fixed point corresponds to the following critical values
\begin{equation} \label{eq:critical_bicon_couplings}
\begin{gathered}
    \alpha = - (d-2)(d-1)/2 + O(N^{-1}),\\  g = 15360  + O(N^{-1}).
\end{gathered}
\end{equation}

\section{Flow to the IR Fixed Point}
Let us couple the critical biconical model \eqref{eq:critical_bicon_model} to $N_f$ massless Dirac fermions $\psi_a$ in the fundamental representation of $U(N_f)$ via a Yukawa interaction with the field~$\chi$,
\begin{equation}\label{eq:model}
\begin{split}
    &S = \int d^dx \bigg\{\frac{1}{2}(\partial \phi)^2 + \frac{1}{2}(\partial \chi)^2 + \bar\psi \slashed{\partial} \psi \\&+ \frac{1}{2\sqrt{N}}\sigma(\phi^2 + \alpha \chi^2)+ \delta_{d,3}\frac{g}{6!N^2}\chi^6 + \frac{y}{\sqrt{M}} \chi \bar\psi \psi \bigg\},
\end{split}
\end{equation}
where $3\leq d<4$ and the Yukawa coupling $y$ has mass dimension $2-d/2$. Here $M=N_f d_\gamma$, with $d_\gamma$ denoting the dimension of the spinor representation. The model has an $O(N)\times U(N_f)\times \mathcal{P}$ symmetry, where~$\mathcal{P}$ is a discrete transformation defined by
\begin{equation}\label{eq:parity}
\mathcal{P}:
x_1\to -x_1,\quad \chi\to -\chi, \quad \psi\to \gamma^1\psi, \quad
\bar\psi\to-\bar\psi\gamma^1.
\end{equation} 
It corresponds to spatial reflection in $d=3$ and is consistent in arbitrary spacetime dimensions, including noninteger $d$~\footnote{In $3+1$ dimensions, it corresponds to the composition of conventional parity and a chiral transformation.}. Although we work in Euclidean signature, where all coordinate directions are equivalent, the distinction between parity $\mathcal P$ and time reversal $\mathcal T$ emerges upon continuation to Lorentzian signature. In $2+1$ dimensions, the fermion bilinear $\bar\psi\psi$ is odd under both symmetries~\cite{Moshe:2003xn,Dunne1999}, so invariance of the Yukawa interaction requires $\chi$ to be odd under both as well.

At $y=0$, the theory is the direct product of the critical biconical model and $N_f$ free massless Dirac fermions. This product CFT serves as the UV fixed point of the RG trajectory triggered by the relevant Yukawa deformation. Sufficiently close to $d=4$, the deformation is weakly relevant, and the flow can be analyzed perturbatively using the $\epsilon$-expansion \cite{SmolkinYung:2025parity}. In contrast, in $d=3$ the Yukawa coupling is strongly relevant, and the model is not analytically tractable without additional assumptions. Consequently, \cite{SmolkinYung:2025parity} employed a numerical truncation of the FRG flow, which does not provide fully controlled error estimates. Here we circumvent this limitation by developing a controlled analytic expansion in $1/N$. 

To this end, we consider $N$ and $M$ large but finite, with their ratio $\nu=M/N$ held fixed. In this regime, the details of the RG flow are not essential, since the IR endpoint can be determined unambiguously by examining how the Yukawa deformation modifies the two-point function of $\chi$. At LO in the $1/N$ expansion, the corresponding self-energy is given by the fermion loop
\begin{center}

\input{pics/cftir_sigma_chi_lo}
\end{center}
where the arrowed line denotes the fermion propagator. Algebraically, this reads
\begin{equation}\label{eq:Sigma_chi}
    \begin{split}
        \Sigma_\chi(p) =\,\, &\frac{y^2}{M} N_f\int_k \Tr \frac{1}{i(\slashed{p}+\slashed{k})} \frac{1}{i\slashed{k}} + O(N^{-1}) \\=\,\, & y^2 \left[-\int_k \frac{1}{k^2}+p^2\Pi(p)\right] + O(N^{-1}),
    \end{split}
\end{equation}
where the trace is taken over spinor indices. With a hard cutoff, the integral $\int_k 1/k^2$ is a pure power-law divergence absorbed into a mass counterterm. Since such divergences are scheme-dependent and, for example, absent in dimensional regularization, we omit them henceforth.
Thus, the inverse propagator, or equivalently the one-particle-irreducible (1PI) two-point vertex, is given by
\begin{equation}
    \Gamma_{\chi\chi}(p) = p^2 + a^{-1} y^2 |p|^{d-2} + O(N^{-1}). 
\end{equation}
As we probe lower momenta, the nonlocal contribution from the fermion loop dominates over the local kinetic term. In the IR, the two-point function scales as $G_{\chi\chi}(p)\sim y^{-2}|p|^{2-d}$, implying that the rescaled field $s=y\chi$ obeys $\Delta_s=1+O(N^{-1})$. The scaling dimension of $\sigma$ remains $\Delta_\sigma=2+O(N^{-1})$. Hence, by large-$N$ factorization, the $(\partial\chi)^2$, $\sigma\chi^2$, and $\chi^6$ terms in~\eqref{eq:model} become irrelevant and drop out in the IR, leading to a fixed-point action of the form
\begin{equation}
    S = \int d^dx \Big\{\frac{1}{2}(\partial \phi)^2 + \frac{1}{2\sqrt{N}}\sigma \phi^2 + \bar\psi \slashed{\partial} \psi + \frac{1}{\sqrt{M}} s \bar\psi \psi \Big\}.
\end{equation}
This action describes a decoupled product CFT: the $(\phi,\sigma)$ sector is the critical $O(N)$ model, while the $(\psi,s)$ sector realizes the renowned Gross--Neveu model \cite{Gross:1974jv} at criticality, written in terms of the Hubbard--Stratonovich field $s$. The critical Gross--Neveu and GNY models are described by the same CFT \cite{Zinn-Justin:1991ksq,Diab:2016spb,Fei:2016sgs,Goykhman:2020ffn}.

We therefore conclude that the relevant Yukawa deformation drives an RG flow from the UV fixed point, given by the product of the critical biconical model and $N_f$ free massless Dirac fermions, to the decoupled product of the critical $O(N)$ and GNY models in the IR~\footnote{At $\alpha=1$ and $g=0$, the same construction also yields an RG flow from the decoupled product of the critical $O(N+1)$ model and $N_f$ free massless Dirac fermions to the decoupled product of the critical $O(N)$ and GNY models.}. The model~\eqref{eq:model} is thus an asymptotically safe quantum field theory with well-defined, nontrivial UV and IR limits.

\section{Persistent Spontaneous Symmetry Breaking}
The model~\eqref{eq:model} is invariant under parity and time reversal. At zero temperature, its infrared physics is governed by the fixed point identified above, namely the decoupled product of the critical $O(N)$ and GNY models. Both have vacua invariant under these symmetries, and therefore the $T=0$ vacuum of~\eqref{eq:model} is as well.

We now turn to the thermal phase structure of~\eqref{eq:model} and ask whether these symmetries remain unbroken at nonzero temperature. We introduce a finite temperature $T$ by compactifying the Euclidean time direction on a circle of circumference $\beta = 1/T$ and imposing periodic and antiperiodic boundary conditions on the bosonic and fermionic fields, respectively. Consequently, the temporal momentum is restricted to the discrete Matsubara frequencies, and the $d$-dimensional Euclidean momentum integrals are replaced by finite-temperature bosonic and fermionic sum-integrals,
\begin{equation*}
\sum_K \equiv T\sum_{n\in\mathbb{Z}}\int\frac{d^{d-1}\mathbf{k}}{(2\pi)^{d-1}}, \quad \sum_{\{K\}}
 \equiv
 T\sum_{n\in\mathbb{Z}+\tfrac{1}{2}}
 \int\frac{d^{d-1}\mathbf{k}}{(2\pi)^{d-1}},
\end{equation*}
where $K = (k_0 , \mathbf{k})$ with $ k_0 = 2\pi n T$.

We now calculate the thermal effective potential and determine its global minima. At LO in the large-$N$ expansion, the effective potential is obtained from the saddle point of the Euclidean functional integral~\cite{PhysRevD.9.1686}.

By an $O(N)$ rotation, we choose $\phi_1\equiv\varphi$ as the only component of the $\phi$ field that may acquire a nonzero expectation value. The remaining $N-1$ transverse components enter quadratically in~\eqref{eq:model} and can therefore be integrated out. For constant $\sigma$, this yields
\begin{equation}
    \frac{N-1}{2}\sum_K\ln\left(K^2+\frac{\sigma}{\sqrt{N}}\right).
\end{equation}
Being Grassmann odd, $\psi_a$ has a vanishing expectation value. 
%The $\psi_a$ field cannot acquire a nonzero expectation value due to rotational invariance. 
Since the fermions enter quadratically, all $N_f$ spinors can be integrated out. For constant $\chi$, this yields the fermionic trace-log
\begin{equation}
\begin{split}
 -N_f\sum_{\{K\}}
 &\Tr\ln\left(i\slashed{K}+\frac{y\chi}{M^{1/2}}\right)\\
 &= 
 -\frac{M}{2}\sum_{\{K\}}
 \ln\left(K^2+(y\chi)^2/M\right).
 \end{split}
 \label{eq:fermionic_thermal_determinant}
\end{equation}
Upon performing the large-$N$ rescaling $\sigma\to \sqrt{N}\sigma$, $\varphi \to \sqrt{N} \varphi$, $\chi \to \sqrt{N} \chi$, the LO effective potential takes the form
\begin{equation}\label{eq:effective_potential_thermal_lo}
\begin{split}
 \frac{1}{N}V_{\mathrm{LO}}(\varphi,\chi,\sigma)
 ={}&
 \frac{1}{2}\sigma(\varphi^2 + \alpha \chi^2)
 +\delta_{d,3}\frac{g}{6!}\chi^6 \\
 +\,&\frac{1}{2}\sum_K\ln\left(K^2+\sigma\right) \\-\,&
 \frac{\nu}{2}\sum_{\{K\}}
 \ln\left(K^2+ y^2\chi^2/\nu\right),
 \end{split}
\end{equation}
where $\nu=M/N$. Although this expression holds for general $d$, we henceforth specialize to the physically relevant $d=3$. In this dimension, we use the LO biconical values $\alpha=-1$ and $g=15360$ given in~\eqref{eq:critical_bicon_couplings}, and evaluate the trace-logs analytically (see the Appendix):
\begin{equation}
\begin{split}
    \frac{1}{2}\sum_{K / \{K\}} &\ln\left(K^2 + m^2\right) = -\frac{m^3}{12\pi} \\ &-\frac{T^3}{2\pi} \left(\frac{m}{T} \text{Li}_2\left(\pm e^{-\frac{m}{T}}\right)+ \text{Li}_3\left(\pm e^{-\frac{m}{T}}\right)\right),
\end{split}
\end{equation}
where $m \geq 0$, and the upper and lower signs correspond to the bosonic and fermionic sum-integrals, respectively. The field $\sigma$ can be eliminated from the effective potential by the saddle-point equation,
\begin{equation} \label{eq:sigma_saddle_eq}
 0= \frac{2}{N}\left(\frac{\partial V_{\mathrm{LO}}}{\partial\sigma}\right)_{\varphi,\chi} = (\varphi^2 + \alpha\chi^2)+\sum_K\frac{1}{K^2+\sigma}, 
\end{equation}
which, upon specializing to $d=3$, admits an analytic solution,
\begin{equation} \label{eq:sigma_saddle}
    \sigma(\varphi,\chi) = 4T^2\,\text{arcsinh}^2\left(\frac{1}{2}e^{2\pi(\varphi^2 - \chi^2)/T}\right).
\end{equation}
The $\varphi$-stationarity condition yields 
\begin{equation}
    0 = \frac{1}{N} \frac{d V_{\text{LO}}(\varphi,\chi, \sigma(\varphi,\chi))}{d \varphi} = \frac{1}{N} \left(\frac{\partial V_{\text{LO}}}{\partial \varphi}\right)_{\chi,\sigma} = \sigma \varphi,
\end{equation}
where the implicit $\varphi$ dependence of $\sigma$ does not contribute because $\partial V_{\text{LO}}/\partial \sigma = 0$ at the saddle point, as imposed by~\eqref{eq:sigma_saddle_eq}. Since $\sigma(\varphi,\chi)$ is strictly positive for all finite field values, the stationarity condition forces the minima of the effective potential to lie on the $\varphi = 0$ locus. This is consistent with the Coleman–Mermin–Wagner theorem, which forbids spontaneous breaking of continuous symmetries at nonzero temperature in two spatial dimensions. We therefore set $\varphi = 0$ and focus on the $\chi$ dependence. The $\chi$-stationarity condition reads
\begin{equation}\label{eq:stationarity}
    \frac{1}{N} \left(\frac{\partial V_{\text{LO}}}{\partial \chi}\right)_\sigma = \chi\left(-\sigma + 128 \chi^4 + \frac{y^2 T}{2\pi} \ln(2\cosh\frac{\tilde m}{2T})\right),
\end{equation}
where $\tilde m = \nu^{-1/2}|y\chi|$. The $\chi=0$ solution is a stationary point. Its stability is determined by the second derivative of the full effective potential at the origin, 
\begin{equation}\label{eq:second_derivative_at_origin}
\begin{split}
    \frac{1}{N} \frac{d^2 V}{d \chi^2}\Bigg|_{\chi = 0} =&\,\, T y^2 \ln 2/(2\pi) -4 T^2 \text{arcsinh}^2\left(1/2\right) \\ +&\,\, O(\ln |t|^{-1}/N),
\end{split}
\end{equation}
where $t=(T_\mathrm{crit}-T)/T_\mathrm{crit}$. In the Appendix, we discuss the logarithmic $\ln|t|^{-1}$ enhancement of the NLO remainder and show that the $1/N$ expansion breaks down for $|t|\lesssim 1/N$. Within its regime of validity, the zero of the second derivative determines the critical temperature,
\begin{equation}\label{eq:critical_temperature}
\begin{split}
    T_{\text{crit}}/y^2 &= \frac{\ln (2)}{8 \pi\,  \text{arcsinh}\left(\frac{1}{2}\right)^2} + O(\ln N/N) \\ &\approx 0.119 + O(\ln N/N).
\end{split}
\end{equation}
Below $T_\mathrm{crit}$, the second derivative is positive, so $\chi=0$ is locally stable. Above $T_\mathrm{crit}$, it becomes negative, and the origin is unstable. Substituting $\sigma$ from~\eqref{eq:sigma_saddle} into~\eqref{eq:stationarity}, we find that additional stationary points satisfy
\begin{equation} \label{eq:nonzero_stationarity_eq}
    4T^2\,\text{arcsinh}^2\left(\frac{1}{2}e^{-2\pi \chi^2/T}\right) = 128 \chi^4 + \frac{y^2 T}{2\pi} \ln(2\cosh\frac{\tilde m}{2T}).
\end{equation}
For $T>T_{\mathrm{crit}}$, this equation admits a pair of locally stable nonzero solutions related by $\chi\to-\chi$. 

The effective potential as a function of $\chi$, with $\varphi$ set to zero and $\sigma$ fixed by~\eqref{eq:sigma_saddle}, reads
\begin{align}\label{eq:effective_potential_chi}
 \frac{1}{N}V_{\mathrm{LO}}(\chi)
 ={}&
 -\frac{1}{2}\sigma\chi^2
 +\frac{64}{3}\chi^6 - \frac{1}{12\pi}\sigma^{3/2} + \frac{\nu^{-1/2}}{12\pi} |y\chi|^3\nonumber\\
 -\,&\frac{T^3}{2\pi} \left(\frac{\sqrt{\sigma}}{T} \text{Li}_2(e^{-\frac{\sqrt{\sigma}}{T}})+ \text{Li}_3(e^{-\frac{\sqrt{\sigma}}{T}})\right) \\
+\, & \frac{\nu T^3}{2 \pi } \left(\frac{\tilde m}{T} \text{Li}_2(-e^{-\frac{\tilde m}{T}})+\text{Li}_3(-e^{-\frac{\tilde m}{T}})\right). \nonumber
\end{align}
The potential is bounded from below, and numerical minimization confirms that the locally stable extrema identified above are its global minima in their respective temperature regimes. Although the effective potential is computed in Euclidean signature, its minima give the corresponding Lorentzian thermal expectation values of $\chi$. The resulting phase structure, shown in Fig.~\ref{fig:chi_tev}, exhibits an inverse thermal transition at $T_{\mathrm{crit}}$, above which the thermal condensate $\langle\chi\rangle_T\neq0$ spontaneously breaks parity and time reversal~\footnote{For even $N_f$, the fermion flavors can be split into two groups of $N_f/2$ with opposite-sign Yukawa couplings,~$y\chi(\bar\psi^L\psi^L-\bar\psi^R\psi^R)$. Parity and time-reversal transformations may then exchange the two groups, allowing $\langle\chi\rangle_T\neq0$ without breaking either symmetry. This alternative model has $U(N_f/2)\times U(N_f/2)$ flavor symmetry and is not considered here. }.
At fixed $T>T_{\mathrm{crit}}$, $\langle\chi\rangle_T^2/T$ increases monotonically with $\nu$, while its high-temperature limit and $T_{\mathrm{crit}}$ are $\nu$-independent at LO.
\afterpage{ %[!t]
\begin{figure}[h]
    \centering
    \includegraphics[width=\linewidth]{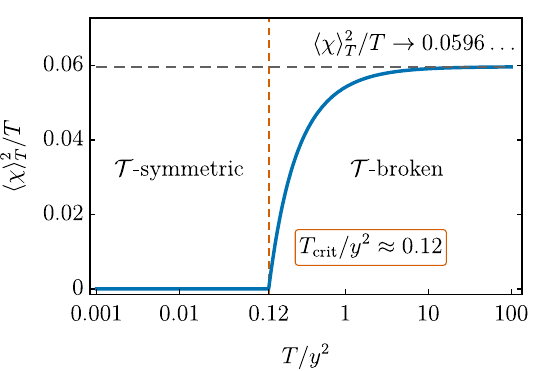}
    \caption{$\langle\chi\rangle_T^2/T$ as a function of $T/y^2$ for $\nu=1$, at LO in $1/N$. The dashed vertical line denotes the critical temperature, above which parity and time reversal are spontaneously broken. The dashed horizontal line indicates the high-temperature biconical limit. }
    \label{fig:chi_tev}
\end{figure}
}

For $|t| \lesssim 1/N$, the $1/N$ expansion breaks down, and the transition is governed by the two-dimensional Ising universality class, as shown in the Appendix. Away from this parametrically narrow window, however, the finite-$N$ corrections remain suppressed: they may shift the transition within an $O(\ln N/N)$ neighborhood but cannot alter the leading-order phase structure for sufficiently large~$N$. The asymptotic high-temperature expectation value is approximately given by
\begin{equation}
    \langle\chi\rangle_T^2/T \xrightarrow[]{T/y^2 \to \infty} 0.0596+O(N^{-1}),
\end{equation}
reproducing the critical biconical limit~\cite{Komargodski:2024zmt,Smolkin:2026wij}. 
The leading term is strictly positive and of order unity, so~$O(N^{-1})$ corrections cannot cancel it for sufficiently large $N$. 
In our construction, a nonzero Yukawa coupling is essential, ensuring that $\chi\to-\chi$ forms part of the spacetime parity transformation~\eqref{eq:parity}, rather than an independent internal $\mathbb{Z}_2$ symmetry.
Thus, for large but finite $N$ and $M$, parity and time reversal are spontaneously broken above a critical temperature, and the broken phase persists to arbitrarily high temperatures.

\section*{Acknowledgments}

We thank Bilal Hawashin and Michael M. Scherer for insightful discussions. This work was supported in part by BSF Grant No.~2022113, ISF Grant No.~2526/25, NSF-BSF Grant No.~2022726, and Israel’s Council for Higher Education.

\bibliography{bibliography}
\clearpage

\appendix   
\setcounter{equation}{0}
\renewcommand{\theequation}{A\arabic{equation}}

\section*{Appendix}
\addcontentsline{toc}{section}{Appendix}

\section{Thermal sum-integrals} \label{app:thermal_sum_integrals}
Consider the following $d=3$ bosonic sum-integral
\begin{equation*}
    \frac{1}{2}\sum_P \ln(P^2 + m^2) \equiv \frac{T}{2} \int \frac{d^2 p}{(2\pi)^2} \sum_n \ln(\omega_n^2 + p^2+m^2),
\end{equation*}
where $m \geq 0$ and $\omega_n = 2\pi n T$. Its $T=0$ part can be obtained by dimensional regularization
\begin{equation}
    \frac{1}{2}\int \frac{d^d p}{(2\pi)^d} \ln(p^2 + m^2)  =- \frac{\Gamma\left(-\frac{d}{2}\right)}{2(4\pi)^{d/2}} m^d.
\end{equation}
To obtain the thermal part, consider first the following Matsubara sum
\begin{equation}
    T \sum_n \frac{1}{\omega_n^2 + E^2} = \frac{1}{2E}\coth{\frac{E}{2T}}.
\end{equation}
Hence, 
\begin{equation}
\begin{split}
    \partial_E \frac{T}{2} \sum_n \ln(\omega_n^2 + E^2) = \frac{1}{2}\coth{\frac{E}{2T}}.
    \end{split}
\end{equation}
Integrating both sides, we obtain
\begin{equation*}
\begin{split}
\frac{T}{2} \sum_n \ln(\omega_n^2 + E^2) = \frac{E}{2} + T\ln\left(1-e^{-E/T}\right) + \text{const}.
    \end{split}
\end{equation*}
We omit the integration constant, as after the spatial-momentum integration, it multiplies the scaleless integral $\int d^2 p$, which vanishes in dimensional regularization. The second term is the thermal part. We integrate it over the spatial momentum with $E=\sqrt{p^2+m^2}$, 
\begin{equation}
    \frac{T}{2\pi} \int_0^\infty pdp \ln\left(1- e^{-\sqrt{p^2+m^2}/T}\right).
\end{equation}
This can be evaluated analytically. Together with the non-thermal part, we obtain
\begin{equation}
\begin{split}
    \frac{1}{2}\sum_P \ln(P^2 +&\, m^2) =  -\frac{m^3}{12\pi} \\[-0.2cm] &-\frac{T^3}{2\pi} \left(\frac{m}{T} \text{Li}_2\left(e^{-\frac{m}{T}}\right)+ \text{Li}_3\left(e^{-\frac{m}{T}}\right)\right).
\end{split}
\end{equation}
In the fermionic case, we consider the following Matsubara sum
\begin{equation*}
    T \sum_n \frac{1}{\tilde\omega_n^2 + E^2} = \frac{1}{2E}\tanh{\frac{E}{2T}}, \quad \tilde \omega_n = 2\pi(n+\tfrac12)T.
\end{equation*}
Performing all the steps analogously to the bosonic calculation, we obtain the fermionic thermal part,
\begin{equation}
    \frac{T}{2\pi} \int_0^\infty pdp \ln\left(1+ e^{-\sqrt{p^2+m^2}/T}\right).
\end{equation}
Therefore, we obtain the following fermionic sum-integral
\begin{equation}
\begin{split}
    \frac{1}{2}\sum_{\{P\}} \ln(&P^2 + m^2) = -\frac{m^3}{12\pi} \\[-0.2cm] & -\frac{T^3}{2\pi} \left(\frac{m}{T} \text{Li}_2\left(-e^{-\frac{m}{T}}\right)+ \text{Li}_3\left(-e^{-\frac{m}{T}}\right)\right).
\end{split}
\end{equation}
Differentiating with respect to $m^2$, we get the following useful identities
\begin{equation}
\begin{split}
    \sum_{P/\{P\}} \frac{1}{P^2 + m^2} = -\frac{m}{4\pi} - \frac{T}{2\pi} \ln\left(1\mp e^{-\frac{m}{T}}\right).
\end{split}
\end{equation}
\section{LO two-point 1PI vertices}
At LO, the thermal effective action coincides with the action obtained by integrating out the $\phi_i$ and $\psi_a$ fields in~\eqref{eq:model}. In $d=3$, it reads
\begin{equation}\label{eq:effective_action_thermal_lo}
\begin{split}
\frac{1}{N}\Gamma_{\mathrm{LO}}[\chi,\sigma]
 = \int d\tau d^2x\Big\{ \frac{1}{2}(\partial \chi)^2 
 -\frac{1}{2}\sigma\chi^2
 +\frac{64}{3}\chi^6 \Big\}\\
 +\frac{1}{2} \Tr \ln\left(-\partial^2+\sigma(x)\right)-
 \frac{\nu}{d_\gamma}\Tr
 \ln\left(\slashed{\partial}+ \frac{y}{\nu^{1/2}} \chi(x)\right).
 \end{split}
\end{equation}
To obtain the LO two-point functions, introduce fluctuations around a constant background $\chi(x) = \chi + \delta\chi(x)$, $\sigma(x) = \sigma + i\delta\sigma(x)$ with constant $\chi$ and $\sigma$. Note that the integration contour of the auxiliary Hubbard–Stratonovich field is parallel to the imaginary axis. The second variation of the LO effective action is then 
\begin{equation}
    \begin{split}
        \mathcal A(K) \equiv \frac{1}{N} &\frac{\delta^2 \Gamma_{\mathrm{LO}}}{\delta \chi(-K) \delta\chi(K)} = K^2 -  \sigma+ 640 \chi^4 \\&+ y^2\left((K^2 + 4\tilde m^2)\tilde\Pi_{\tilde m^2}(K)-\sum_{\{Q\}} \frac{1}{Q^2+\tilde m^2}\right),\nonumber
    \end{split}
\end{equation}~
\begin{align}
    \frac{1}{N}&\frac{\delta^2 \Gamma_{\mathrm{LO}}}{\delta \sigma(-K) \delta\sigma(K)} = \Pi_{\sigma}(K),\\ \frac{1}{N}&\frac{\delta^2 \Gamma_{\mathrm{LO}}}{\delta \chi(-K) \delta\sigma(K)} = \frac{1}{N}\frac{\delta^2 \Gamma_{\mathrm{LO}}}{\delta \sigma(-K) \delta\chi(K)} = - i \chi,\nonumber
\end{align}
where $\tilde m = \nu^{-1/2}|y\chi|$, and
\begin{equation}
\begin{split}
    \Pi_{m^2}(K) = \frac{1}{2}\sum_P \frac{1}{P^2 + m^2}\frac{1}{(K+P)^2+m^2}, \\\tilde \Pi_{m^2}(K) = \frac{1}{2}\sum_{\{P\}} \frac{1}{P^2 + m^2}\frac{1}{(K+P)^2+m^2}.
    \end{split}
\end{equation}
The LO $\chi$ propagator reads
\begin{equation}
    G_{\chi\chi}^\mathrm{LO}(K) = \frac{1}{N} \frac{\Pi_\sigma (K)}{D(K)}, \quad D(K) = \mathcal A(K) \Pi_\sigma(K)+\chi^2,
\end{equation}
where $\sigma$ takes its constant saddle-point value~\eqref{eq:sigma_saddle}.
Consider the soft momentum region with zero Matsubara mode $K = (0, \mathbf{k})$. Using the Feynman parametrization, we obtain the bosonic and fermionic bubble expansions 
\begin{widetext}
\begin{equation}
    \begin{split}
            \Pi_{m^2} (\mathbf{k}) = &\,\frac{1}{2} \sum_P \int_0^1 dx \frac{1}{[P^2 + m^2 +x(1-x)\mathbf{k}^2]^2}  = \frac{1}{2}\sum_P \frac{1}{(P^2 + m^2)^2} - \frac{\mathbf{k}^2}{6} \sum_P \frac{1}{(P^2 + m^2)^3} + O(\mathbf{k}^4)\\=&\,\frac{\coth \left(\frac{m}{2 T}\right)}{16 \pi  m} -\frac{\mathbf{k}^2}{192 \pi  m^3}\left(\coth \left(\frac{m}{2 T}\right)+\frac{m}{2 T} \text{csch}^2\left(\frac{m}{2 T}\right)\right) + O(\mathbf{k}^4),\\[0.2cm]
            \tilde \Pi_{\tilde m^2} (\mathbf{k}) =&\, \frac{\tanh \left(\frac{\tilde m}{2 T}\right)}{16 \pi  \tilde m} -\frac{\mathbf{k}^2}{192 \pi \tilde m^3} \left(\tanh \left(\frac{\tilde m}{2 T}\right)-\frac{\tilde m}{2 T} \text{sech}^2\left(\frac{\tilde m}{2 T}\right)\right) + O(\mathbf{k}^4).
    \end{split}
\end{equation}
\end{widetext}
In the following, we analyze the vicinity of the phase transition, namely temperatures 
$$|t_\mathrm{LO}|\ll1, \qquad t_\mathrm{LO}=(T_\mathrm{crit}^\mathrm{LO}-T)/T_\mathrm{crit}^\mathrm{LO}.$$
For simplicity, we take $t_\mathrm{LO}\geq 0$ when calculating correlators, so that the corresponding background field minimum is at~$\chi = 0$. In the soft momentum region at zero Matsubara mode, the $\chi$ propagator reads
\begin{align}
    G_{\chi\chi}^{\mathrm{LO}}(\mathbf{k}) &= \frac{1}{N} \frac{\Pi_\sigma (\mathbf{k})}{D(\mathbf{k})} \\&= \frac{1}{N} \frac{1}{c_1\,\mathbf{k}^2+c_2\,y^4 t_\mathrm{LO} + O(\mathbf{k}^4/y^4, y^4 t_\mathrm{LO}^2)},\nonumber
\end{align}
where $c_i>0$ are order-one coefficients. The 1PI $4$-point $\chi$ vertex at zero momentum reads
\begin{equation}
\begin{split}
    \Gamma_{\chi}^{(4)}(K_i=0) =\,& \frac{d^4 V_{\text{LO}}(\chi,\sigma(\chi))}{d\chi^4} + O(N^0)\\ =\,& c_3\,y^2 N + O(N^0).
\end{split}
\end{equation}
One contribution to the NLO correction to the $\chi$ mass is the one-loop tadpole: two legs of the quartic vertex are external, while the other two contract with each other, so there is one internal propagator. We consider the soft momentum contribution:
\begin{equation*}
\begin{split}
    &\delta \Gamma^{(2)}_{\chi\chi} = \frac{1}{2}\sum_K\Gamma_{\chi}^{(4)}G_{\chi\chi} + \ldots \\&\stackrel{\mathrm{soft}}{\approx} (y^2 N)\, T \int \frac{d^2 \mathbf{k}}{(2\pi)^2} \frac{1}{N}\frac{1}{c_1\mathbf{k}^2+c_2y^4 t_\mathrm{LO}} \sim y^4 \ln 1/t_\mathrm{LO},
\end{split}
\end{equation*}
where $T\simeq T_\mathrm{crit}=O(y^2)$. This contribution controls the NLO remainder in the second derivative of the effective potential at the origin~\eqref{eq:second_derivative_at_origin}. 
For $t_\mathrm{LO}\sim \ln N/N$, it becomes comparable to $\Gamma^{(2)}_{\mathrm{LO},\chi\chi}\sim N y^4 t_\mathrm{LO}$ and hence $\ln N/N$ controls the NLO displacement of the critical temperature.

\,\\[-1.2cm]
\subsection*{Breakdown of the $1/N$ expansion near the phase transition} %at $T\simeq T_\mathrm{crit}$
Subsequent $1/N$ corrections to the effective potential~\eqref{eq:effective_potential_chi} are obtained by computing vacuum diagrams at constant background fields~\cite{Coleman:1985rnk,Smolkin:2026wij}. One class of such diagrams is the $\chi$ loops connected via $4$-point interactions. The large-$N$ saddle point admits a loop expansion, so increasing the number of loops in a diagram corresponds to a higher $1/N$ order contribution to the effective potential. However, suppose we add a loop to a vacuum diagram with $n+1$ loops that contributes at order $1/N^{n}$. Then the relative factor integrated in the soft momentum region reads
\begin{align}
    \sum_K \Gamma^{(4)}_\chi G_{\chi\chi}^2 &\stackrel{\mathrm{soft}}{\approx} (y^2 N)\, T \int_{k\lesssim \mathbf{k}_*} \frac{d^2 \mathbf{k}}{(2\pi)^2} \frac{1}{N^2}\frac{1}{(c_1\mathbf{k}^2+ c_2y^4 t)^2} \nonumber \\&\,\,\sim O\left(\frac{1}{Nt}\right).
\end{align}
where $t$ is defined relative to the exact critical temperature and taken positive, while $\mathbf{k}_* \sim y^2t^{1/2}$ is the soft momentum scale below which $G_{\chi\chi}(\mathbf{k}) \sim 1/(y^4Nt)$. This computation shows that in the near-critical region $t\lesssim 1/N$, where the $\chi$ field becomes nearly massless, successive soft-loop corrections to the effective potential become comparable to or larger than the preceding terms: $V^{(n+1)}_\mathrm{soft}\gtrsim V^{(n)}_\mathrm{soft}$. This manifests the breakdown of the $1/N$ expansion in the vicinity of the phase transition $|t|\lesssim 1/N$.
\\[-1.1cm]
\section{The phase transition} 
At $T=T_\mathrm{crit}$, the Matsubara zero mode of $\chi$ is the only degree of freedom that becomes massless. All other modes and fields remain gapped: the saddle-point value of $\sigma$ is the squared mass of the $\phi_i$ fields, while the fermions do not have a zero Matsubara mode, so along with the nonzero Matsubara $\chi$ modes they have a gap~$\sim T$. Integrating out the gapped modes yields an effective local two-dimensional theory of a single scalar. Expanding the action~\eqref{eq:effective_action_thermal_lo} at small $\chi_0$ and spatial momentum, and using the preceding results, gives 
\begin{equation}
\begin{split}
    S_\mathrm{2d}^\mathrm{LO} = \frac{N}{T} \int d^2x\bigg\{\frac{c_1}{2}(\partial \chi_0)^2+&\frac{1}{2}(c_2y^4t_\mathrm{LO}) \chi_0^2\\+&\frac{1}{4!}(c_3y^2)\chi_0^4+ \ldots\bigg\}.
\end{split}
\end{equation}
Further loops of the gapped fluctuations generate only infrared-finite $1/N$ corrections. The quartic coupling in the reduced action is positive, while varying $T$ tunes its quadratic coupling through the critical surface. The resulting critical theory therefore belongs to the two-dimensional Ising universality class~\cite{Yang:1952xj}.

\end{document}

%% file: pics/bubble_chain.tex
\tikzset{every picture/.style={line width=0.75pt}} %set default line width to 0.75pt        

\begin{tikzpicture}[x=0.75pt,y=0.75pt,yscale=-1,xscale=1]
%uncomment if require: \path (0,102); %set diagram left start at 0, and has height of 102

%Straight Lines [id:da7022214467051172] 
\draw  [dash pattern={on 2.25pt off 2.25pt}]  (128.77,45.34) -- (105.3,45.34) ;
%Curve Lines [id:da5733355096463146] 
\draw    (169.01,45.59) .. controls (170.04,33.03) and (185.79,35.08) .. (185.76,45.59) ;
\draw [shift={(185.76,45.59)}, rotate = 90.16] [color={rgb, 255:red, 0; green, 0; blue, 0 }  ][fill={rgb, 255:red, 0; green, 0; blue, 0 }  ][line width=0.75]      (0, 0) circle [x radius= 1.34, y radius= 1.34]   ;
\draw [shift={(169.01,45.59)}, rotate = 274.71] [color={rgb, 255:red, 0; green, 0; blue, 0 }  ][fill={rgb, 255:red, 0; green, 0; blue, 0 }  ][line width=0.75]      (0, 0) circle [x radius= 1.34, y radius= 1.34]   ;
%Curve Lines [id:da48856455753435557] 
\draw    (185.76,45.59) .. controls (184.73,58.15) and (168.98,56.09) .. (169.01,45.59) ;
%Curve Lines [id:da07237044048694496] 
\draw    (239.01,45.59) .. controls (240.04,33.03) and (255.79,35.08) .. (255.76,45.59) ;
\draw [shift={(255.76,45.59)}, rotate = 90.16] [color={rgb, 255:red, 0; green, 0; blue, 0 }  ][fill={rgb, 255:red, 0; green, 0; blue, 0 }  ][line width=0.75]      (0, 0) circle [x radius= 1.34, y radius= 1.34]   ;
\draw [shift={(239.01,45.59)}, rotate = 274.71] [color={rgb, 255:red, 0; green, 0; blue, 0 }  ][fill={rgb, 255:red, 0; green, 0; blue, 0 }  ][line width=0.75]      (0, 0) circle [x radius= 1.34, y radius= 1.34]   ;
%Curve Lines [id:da008330827444802846] 
\draw    (255.76,45.59) .. controls (254.73,58.15) and (238.98,56.09) .. (239.01,45.59) ;
%Curve Lines [id:da19457118230618398] 
\draw    (270,45.59) .. controls (271.03,33.03) and (286.78,35.08) .. (286.75,45.59) ;
\draw [shift={(286.75,45.59)}, rotate = 90.16] [color={rgb, 255:red, 0; green, 0; blue, 0 }  ][fill={rgb, 255:red, 0; green, 0; blue, 0 }  ][line width=0.75]      (0, 0) circle [x radius= 1.34, y radius= 1.34]   ;
\draw [shift={(270,45.59)}, rotate = 274.71] [color={rgb, 255:red, 0; green, 0; blue, 0 }  ][fill={rgb, 255:red, 0; green, 0; blue, 0 }  ][line width=0.75]      (0, 0) circle [x radius= 1.34, y radius= 1.34]   ;
%Curve Lines [id:da11070771438712768] 
\draw    (286.75,45.59) .. controls (285.71,58.15) and (269.97,56.09) .. (270,45.59) ;
%Shape: Boxed Line [id:dp927262805391632] 
\draw  [dash pattern={on 2.25pt off 2.25pt}]  (154.77,45.65) -- (169.01,45.52) ;
%Straight Lines [id:da04551465713199032] 
\draw  [dash pattern={on 2.25pt off 2.25pt}]  (200,45.59) -- (185.76,45.59) ;
%Shape: Boxed Line [id:dp7251121969877095] 
\draw  [dash pattern={on 2.25pt off 2.25pt}]  (224.77,45.54) -- (239.01,45.64) ;
%Straight Lines [id:da7690366673168967] 
\draw  [dash pattern={on 2.25pt off 2.25pt}]  (270,45.59) -- (255.76,45.59) ;
%Straight Lines [id:da404313446913377] 
\draw  [dash pattern={on 2.25pt off 2.25pt}]  (300.99,45.59) -- (286.75,45.59) ;

% Text Node
\draw (135,39.07) node [anchor=north west][inner sep=0.75pt]    {$+$};
% Text Node
\draw (205,39.07) node [anchor=north west][inner sep=0.75pt]    {$+$};
% Text Node
\draw (135,63.07) node [anchor=north west][inner sep=0.75pt]    {$=-\lambda _{\phi } +\lambda _{\phi }^{2} \Pi ( p) -\lambda _{\phi }^{3} \Pi ^{2}( p) +\dotsc =-\frac{\lambda _{\phi }}{1+\lambda _{\phi } \Pi ( p)} ,$};
% Text Node
\draw (309,39.07) node [anchor=north west][inner sep=0.75pt]    {$+\ \dotsc $};

\end{tikzpicture}

%% file: pics/cftir_sigma_chi_lo.tex
\tikzset{every picture/.style={line width=0.75pt}} %set default line width to 0.75pt        

\begin{tikzpicture}[x=0.75pt,y=0.75pt,yscale=-1,xscale=1]
%uncomment if require: \path (0,76); %set diagram left start at 0, and has height of 76

%Straight Lines [id:da4946503907222428] 
\draw    (282.1,34.57) -- (305.01,34.57) ;
\draw [shift={(305.01,34.57)}, rotate = 0] [color={rgb, 255:red, 0; green, 0; blue, 0 }  ][fill={rgb, 255:red, 0; green, 0; blue, 0 }  ][line width=0.75]      (0, 0) circle [x radius= 1.34, y radius= 1.34]   ;
%Straight Lines [id:da15619128229848345] 
\draw    (337.14,34.19) -- (360.05,34.19) ;
\draw [shift={(337.14,34.19)}, rotate = 0] [color={rgb, 255:red, 0; green, 0; blue, 0 }  ][fill={rgb, 255:red, 0; green, 0; blue, 0 }  ][line width=0.75]      (0, 0) circle [x radius= 1.34, y radius= 1.34]   ;
%Curve Lines [id:da22737172199709776] 
\draw    (305.01,34.57) .. controls (306.99,10.48) and (337.19,14.42) .. (337.13,34.57) ;
\draw [shift={(325.53,18.63)}, rotate = 180] [fill={rgb, 255:red, 0; green, 0; blue, 0 }  ][line width=0.08]  [draw opacity=0] (8.04,-3.86) -- (0,0) -- (8.04,3.86) -- cycle    ;
%Curve Lines [id:da17842832784123908] 
\draw    (337.13,34.57) .. controls (335.15,58.65) and (304.96,54.72) .. (305.01,34.57) ;
\draw [shift={(316.62,50.5)}, rotate = 6.01] [fill={rgb, 255:red, 0; green, 0; blue, 0 }  ][line width=0.08]  [draw opacity=0] (8.04,-3.86) -- (0,0) -- (8.04,3.86) -- cycle    ;

% Text Node
\draw (218,26) node [anchor=north west][inner sep=0.75pt]    {$\Sigma _{\chi }( p) =$};
% Text Node
\draw (369.2,23.39) node [anchor=north west][inner sep=0.75pt]    {$+\ O\left( N^{-1}\right) ,$};
\end{tikzpicture}

%% file: bibliography.bib
@article{weinberg1974gauge,
  author  = {Steven Weinberg},
  title   = {Gauge and Global Symmetries at High Temperature},
  journal = {Physical Review D},
  volume  = {9},
  number  = {12},
  pages   = {3357--3378},
  year    = {1974},
  doi     = {10.1103/PhysRevD.9.3357}
}

@article{PhysRevD.9.1686,
  title = {Functional evaluation of the effective potential},
  author = {Jackiw, R.},
  journal = {Phys. Rev. D},
  volume = {9},
  issue = {6},
  pages = {1686--1701},
  numpages = {0},
  year = {1974},
  month = {Mar},
  publisher = {American Physical Society},
  doi = {10.1103/PhysRevD.9.1686},
  url = {https://link.aps.org/doi/10.1103/PhysRevD.9.1686}
}

@article{Chai:2020zgq,
    author = "Chai, Noam and Chaudhuri, Soumyadeep and Choi, Changha and Komargodski, Zohar and Rabinovici, Eliezer and Smolkin, Michael",
    title = "{Thermal Order in Conformal Theories}",
    eprint = "2005.03676",
    archivePrefix = "arXiv",
    primaryClass = "hep-th",
    doi = "10.1103/PhysRevD.102.065014",
    journal = "Phys. Rev. D",
    volume = "102",
    number = "6",
    pages = "065014",
    year = "2020"
}

@article{Chai:2020onq,
    author = "Chai, Noam and Chaudhuri, Soumyadeep and Choi, Changha and Komargodski, Zohar and Rabinovici, Eliezer and Smolkin, Michael",
    title = "{Symmetry Breaking at All Temperatures}",
    doi = "10.1103/PhysRevLett.125.131603",
    journal = "Phys. Rev. Lett.",
    volume = "125",
    number = "13",
    pages = "131603",
    year = "2020"
}

@article{Chai:2021djc,
    author = "Chai, Noam and Dymarsky, Anatoly and Smolkin, Michael",
    title = "{Model of Persistent Breaking of Discrete Symmetry}",
    eprint = "2106.09723",
    archivePrefix = "arXiv",
    primaryClass = "hep-th",
    doi = "10.1103/PhysRevLett.128.011601",
    journal = "Phys. Rev. Lett.",
    volume = "128",
    number = "1",
    pages = "011601",
    year = "2022"
}

@article{Chai:2021tpt,
    author = "Chai, Noam and Dymarsky, Anatoly and Goykhman, Mikhail and Sinha, Ritam and Smolkin, Michael",
    title = "{A model of persistent breaking of continuous symmetry}",
    eprint = "2111.02474",
    archivePrefix = "arXiv",
    primaryClass = "hep-th",
    doi = "10.21468/SciPostPhys.12.6.181",
    journal = "SciPost Phys.",
    volume = "12",
    number = "6",
    pages = "181",
    year = "2022"
}

@article{Liendo:2022bmv,
    author = "Liendo, Pedro and Rong, Junchen and Zhang, Haoyu",
    title = "{Spontaneous breaking of finite group symmetries at all temperatures}",
    eprint = "2205.13964",
    archivePrefix = "arXiv",
    primaryClass = "hep-th",
    doi = "10.21468/SciPostPhys.14.6.168",
    journal = "SciPost Phys.",
    volume = "14",
    number = "6",
    pages = "168",
    year = "2023"
}

@article{Hawashin:2024dpp,
  title = {Ultraviolet-Complete Local Field Theory of Persistent Symmetry Breaking in $2+1$ Dimensions},
  author = {Hawashin, Bilal and Rong, Junchen and Scherer, Michael M.},
  journal = {Phys. Rev. Lett.},
  volume = {134},
  issue = {4},
  pages = {041602},
  numpages = {6},
  year = {2025},
  month = {Jan},
  publisher = {American Physical Society},
  doi = {10.1103/PhysRevLett.134.041602},
  url = {https://link.aps.org/doi/10.1103/PhysRevLett.134.041602}
}

@article{Komargodski:2024zmt,
    author = "Komargodski, Zohar and Popov, Fedor K.",
    title = "{Temperature-Resistant Order in 2+1 Dimensions}",
    eprint = "2412.09459",
    archivePrefix = "arXiv",
    primaryClass = "hep-th",
    month = "12",
    year = "2024",
    journal=""
}

@article{Han:2025eiw,
    author = "Han, Yiqiu and Huang, Xiaoyang and Komargodski, Zohar and Lucas, Andrew and Popov, Fedor K.",
    title = "{Entropic Order}",
    eprint = "2503.22789",
    archivePrefix = "arXiv",
    primaryClass = "cond-mat.stat-mech",
    month = "3",
    year = "2025",
    journal=""
}

@article{Chaudhuri:2026ges,
    author = "Chaudhuri, Soumyadeep and Hawashin, Bilal and Rabinovici, Eliezer and Scherer, Michael M.",
    title = "{Exploring thermal order in conformal theories with multiple scalars coupled to an $O(N)$ vector field}",
    journal = {},
    eprint = "2609.00160",
    archivePrefix = "arXiv",
    primaryClass = "hep-th",
    reportNumber = "CERN-TH-2026-205",
    month = "8",
    year = "2026"
}

@article{Chaudhuri:2020xxb,
    author = "Chaudhuri, Soumyadeep and Choi, Changha and Rabinovici, Eliezer",
    title = "{Thermal order in large N conformal gauge theories}",
    eprint = "2011.13981",
    archivePrefix = "arXiv",
    primaryClass = "hep-th",
    doi = "10.1007/JHEP04(2021)203",
    journal = "{JHEP}",
    volume = "04",
    pages = "203",
    year = "2021"
}

@article{Bajc:2020gpa,
    author = "Bajc, Borut and Lugo, Adri{\'a}n and Sannino, Francesco",
    title = "{Asymptotically free and safe fate of symmetry nonrestoration}",
    eprint = "2012.08428",
    archivePrefix = "arXiv",
    primaryClass = "hep-th",
    doi = "10.1103/PhysRevD.103.096014",
    journal = "Phys. Rev. D",
    volume = "103",
    pages = "096014",
    year = "2021"
}

@article{Chaudhuri:2021dsq,
    author = "Chaudhuri, Soumyadeep and Rabinovici, Eliezer",
    title = "{Symmetry breaking at high temperatures in large N gauge theories}",
    eprint = "2106.11323",
    archivePrefix = "arXiv",
    primaryClass = "hep-th",
    doi = "10.1007/JHEP08(2021)148",
    journal = "{JHEP}",
    volume = "08",
    pages = "148",
    year = "2021"
}

@article{Buchel:2020thm,
    author = "Buchel, Alex",
    title = "{Thermal order in holographic CFTs and no-hair theorem violation in black branes}",
    eprint = "2005.07833",
    archivePrefix = "arXiv",
    primaryClass = "hep-th",
    doi = "10.1016/j.nuclphysb.2021.115425",
    journal = "Nucl. Phys. B",
    volume = "967",
    pages = "115425",
    year = "2021"
}

@article{Buchel:2020jfs,
    author = "Buchel, Alex",
    title = "{Fate of the conformal order}",
    eprint = "2011.11509",
    archivePrefix = "arXiv",
    primaryClass = "hep-th",
    doi = "10.1103/PhysRevD.103.026008",
    journal = "Phys. Rev. D",
    volume = "103",
    number = "2",
    pages = "026008",
    year = "2021"
}

@article{Buchel:2021ead,
    author = "Buchel, Alex",
    title = "{Compactified holographic conformal order}",
    eprint = "2107.05086",
    archivePrefix = "arXiv",
    primaryClass = "hep-th",
    doi = "10.1016/j.nuclphysb.2021.115605",
    journal = "Nucl. Phys. B",
    volume = "973",
    pages = "115605",
    year = "2021"
}

@article{Buchel:2022zxl,
    author = "Buchel, Alex",
    title = "{The quest for a conifold conformal order}",
    eprint = "2205.00612",
    archivePrefix = "arXiv",
    primaryClass = "hep-th",
    doi = "10.1007/JHEP08(2022)080",
    journal = "{JHEP}",
    volume = "08",
    pages = "080",
    year = "2022"
}

@article{Buchel:2023zpe,
    author = "Buchel, Alex",
    title = "{Holographic conformal order with higher derivatives}",
    eprint = "2312.15764",
    archivePrefix = "arXiv",
    primaryClass = "hep-th",
    doi = "10.1016/j.nuclphysb.2024.116578",
    journal = "Nucl. Phys. B",
    volume = "1004",
    pages = "116578",
    year = "2024"
}

@article{Buchel:2025cve,
    author = "Buchel, Alex",
    title = "{The ordered phase of charged N=4 SYM plasma}",
    eprint = "2501.01856",
    archivePrefix = "arXiv",
    primaryClass = "hep-th",
    month = "1",
    year = "2025",
    journal=""
}

@article{Buchel:2025jup,
    author = "Buchel, Alex",
    title = "{Instability of baryonic black branes}",
    eprint = "2502.05971",
    archivePrefix = "arXiv",
    primaryClass = "hep-th",
    doi = "10.1007/JHEP05(2025)215",
    journal = "{JHEP}",
    volume = "05",
    pages = "215",
    year = "2025"
}

@article{Moshe:2003xn,
    author = "Moshe, Moshe and Zinn-Justin, Jean",
    title = "{Quantum field theory in the large N limit: A Review}",
    eprint = "hep-th/0306133",
    archivePrefix = "arXiv",
    doi = "10.1016/S0370-1573(03)00263-1",
    journal = "Phys. Rept.",
    volume = "385",
    pages = "69--228",
    year = "2003"
}

@article{Fisher_mcl:1974,
  title = {Renormalization-Group Analysis of Bicritical and Tetracritical Points},
  author = {Nelson, David R. and Kosterlitz, J. M. and Fisher, Michael E.},
  journal = {Phys. Rev. Lett.},
  volume = {33},
  issue = {14},
  pages = {813--817},
  numpages = {0},
  year = {1974},
  month = {Sep},
  publisher = {American Physical Society},
  doi = {10.1103/PhysRevLett.33.813},
  url = {https://link.aps.org/doi/10.1103/PhysRevLett.33.813}
}

@article{Vicari:2003,
  title = {Multicritical phenomena in $\mathrm{O}{(n}_{1})\ensuremath{\bigoplus}\mathrm{O}{(n}_{2})$-symmetric theories},
  author = {Calabrese, Pasquale and Pelissetto, Andrea and Vicari, Ettore},
  journal = {Phys. Rev. B},
  volume = {67},
  issue = {5},
  pages = {054505},
  numpages = {12},
  year = {2003},
  month = {Feb},
  publisher = {American Physical Society},
  doi = {10.1103/PhysRevB.67.054505},
  url = {https://link.aps.org/doi/10.1103/PhysRevB.67.054505}
}

@book{Coleman:1985rnk,
    author = "Coleman, Sidney",
    title = "{Aspects of Symmetry}: {Selected Erice Lectures}",
    doi = "10.1017/CBO9780511565045",
    isbn = "978-0-521-31827-3",
    publisher = "Cambridge University Press",
    address = "Cambridge, U.K.",
    year = "1985"
}

@article{SmolkinYung:2025parity,
    author = "Hawashin, Bilal and Scherer, Michael M. and Smolkin, Michael and Yung, Lev",
    title = "{Spontaneous Space-Time Parity Breaking Without Thermal Restoration}",
    journal = {},
    eprint = "2507.19890",
    archivePrefix = "arXiv",
    primaryClass = "hep-th",
    month = "7",
    year = "2025"
}

@article{Bajc:2026ppk,
    author = "Bajc, Borut and Muco, Giulia and Sannino, Francesco and Wagner, Sophie",
    title = "{Infinite heat order in 3+1 dimensions}",
    eprint = "2604.01184",
    archivePrefix = "arXiv",
    primaryClass = "hep-th",
    doi = "10.1103/tfz8-yf7t",
    journal = "Phys. Rev. D",
    volume = "113",
    number = "12",
    pages = "125029",
    year = "2026"
}

@article{Huang:2025gvi,
    author = "Huang, Xiaoyang and Komargodski, Zohar and Lucas, Andrew and Popov, Fedor K. and Sulejmanpasic, Tin",
    title = "{Minimal Models of Entropic Order}",
    journal = {},
    eprint = "2512.07980",
    archivePrefix = "arXiv",
    primaryClass = "cond-mat.stat-mech",
    month = "12",
    year = "2025"
}

@article{Andriolo:2026udg,
    author = "Andriolo, Enrico and Nguyen, Mendel and Richards, Emily and Sulejmanpasic, Tin",
    title = "{Proof of entropic order in Generalized Ising Models}",
    journal = {},
    eprint = "2604.09768",
    archivePrefix = "arXiv",
    primaryClass = "cond-mat.stat-mech",
    month = "4",
    year = "2026"
}

@article{Smolkin:2026wij,
    author = "Smolkin, Michael and Yung, Lev",
    title = "{Thermal Order in the Biconical Model}",
    eprint = "2608.02720",
    archivePrefix = "arXiv",
    primaryClass = "hep-th",
    journal = {},
    month = "8",
    year = "2026"
}

@article{Yang:1952xj,
    author = "Yang, C. N.",
    title = "{The Spontaneous Magnetization of a Two-Dimensional Ising Model}",
    doi = "10.1103/PhysRev.85.808",
    journal = "Phys. Rev.",
    volume = "85",
    pages = "808--816",
    year = "1952"
}

@book{Weinberg1995,
  author    = {Steven Weinberg},
  title     = {The Quantum Theory of Fields},
  volume    = {1},
  subtitle  = {Foundations},
  publisher = {Cambridge University Press},
  year      = {1995},
  doi       = {10.1017/CBO9781139644167}
}

@article{Lee:1956qn,
    author = "Lee, T. D. and Yang, Chen-Ning",
    title = "{Question of Parity Conservation in Weak Interactions}",
    doi = "10.1103/PhysRev.104.254",
    journal = "Phys. Rev.",
    volume = "104",
    pages = "254--258",
    year = "1956"
}

@article{Wu:1957my,
    author = "Wu, C. S. and Ambler, E. and Hayward, R. W. and Hoppes, D. D. and Hudson, R. P.",
    title = "{Experimental Test of Parity Conservation in $\beta$ Decay}",
    doi = "10.1103/PhysRev.105.1413",
    journal = "Phys. Rev.",
    volume = "105",
    pages = "1413--1414",
    year = "1957"
}

@article{Christenson:1964fg,
    author = "Christenson, J. H. and Cronin, J. W. and Fitch, V. L. and Turlay, R.",
    title = "{Evidence for the $2\pi$ Decay of the $K_2^0$ Meson}",
    doi = "10.1103/PhysRevLett.13.138",
    journal = "Phys. Rev. Lett.",
    volume = "13",
    pages = "138--140",
    year = "1964"
}

@article{Haldane:1988,
    author  = {Haldane, F. D. M.},
    title   = "{Model for a Quantum Hall Effect without Landau Levels: Condensed-Matter Realization of the ``Parity Anomaly''}",
    journal = {Phys. Rev. Lett.},
    volume  = {61},
    pages   = {2015--2018},
    year    = {1988},
    doi     = {10.1103/PhysRevLett.61.2015}
}

@article{Nagaosa:2009ycg,
    author = "Nagaosa, Naoto and Sinova, Jairo and Onoda, Shigeki and MacDonald, A. H. and Ong, N. P.",
    title = "{Anomalous Hall effect}",
    eprint = "0904.4154",
    archivePrefix = "arXiv",
    primaryClass = "cond-mat.mes-hall",
    doi = "10.1103/RevModPhys.82.1539",
    journal = "Rev. Mod. Phys.",
    volume = "82",
    number = "2",
    pages = "1539--1592",
    year = "2010"
}

@article{Chang:2023,
    author  = {Chang, Cui-Zu and Liu, Chao-Xing and MacDonald, Allan H.},
    title   = "{Colloquium: Quantum anomalous Hall effect}",
    journal = {Rev. Mod. Phys.},
    volume  = {95},
    pages   = {011002},
    year    = {2023},
    doi     = {10.1103/RevModPhys.95.011002}
}

@article{Fujimoto:1984hr,
    author = "Fujimoto, Yasushi and Sakakibara, Susumu",
    title = "{On symmetry non-restoration at high temperature}",
    reportNumber = "RIFP-572",
    doi = "10.1016/0370-2693(85)90847-0",
    journal = "Phys. Lett. B",
    volume = "151",
    pages = "260--262",
    year = "1985"
}

@article{Salomonson:1984px,
    author = "Salomonson, Per and Skagerstam, Bo-Sture K.",
    title = "{High Temperature Phases in an O($N$) X O($N$) Symmetric Four $\epsilon$ Dimensional Vector Model}",
    reportNumber = "NORDITA-84/30",
    doi = "10.1016/0370-2693(85)91039-1",
    journal = "Phys. Lett. B",
    volume = "155",
    pages = "100--102",
    year = "1985"
}

@article{Klimenko:1988ng,
    author = "Klimenko, K. G.",
    title = "{1/N expansion in the O(N) x O(N) scalar theory and the problem of symmetry restoration at high temperature}",
    reportNumber = "IFVE-88-149, IFVE-87-166",
    doi = "10.1007/BF01016185",
    journal = "Theor. Math. Phys.",
    volume = "80",
    pages = "929--935",
    year = "1989"
}

@article{Grabowski:1990qc,
    author = "Grabowski, Marek P.",
    title = "{The Effective potential and symmetry breaking in the O(N) x O(N) model}",
    reportNumber = "VPI-IHEP-90-3",
    doi = "10.1007/BF01572032",
    journal = "Z. Phys. C",
    volume = "48",
    pages = "505--510",
    year = "1990"
}

@article{Roos:1995vm,
    author = "Roos, Thomas G.",
    title = "{Wilson renormalization group study of inverse symmetry breaking}",
    eprint = "hep-th/9511073",
    archivePrefix = "arXiv",
    reportNumber = "CLNS-95-1373",
    doi = "10.1103/PhysRevD.54.2944",
    journal = "Phys. Rev. D",
    volume = "54",
    pages = "2944--2959",
    year = "1996"
}

@article{Orloff:1996yn,
    author = "Orloff, Jean",
    title = "{The UV price for symmetry nonrestoration}",
    eprint = "hep-ph/9611398",
    archivePrefix = "arXiv",
    reportNumber = "ENSLAPP-AL-615-96",
    doi = "10.1016/S0370-2693(97)00552-2",
    journal = "Phys. Lett. B",
    volume = "403",
    pages = "309--315",
    year = "1997"
}

@article{Gavela:1998ux,
    author = "Gavela, M. B. and Pene, O. and Rius, N. and Vargas-Castrillon, S.",
    title = "{The Fading of symmetry nonrestoration at finite temperature}",
    eprint = "hep-ph/9801244",
    archivePrefix = "arXiv",
    reportNumber = "LPTHE-ORSAY-96-36A, FTUAM-FEV-96-22, FTUV-97-73, IFIC-97-105",
    doi = "10.1103/PhysRevD.59.025008",
    journal = "Phys. Rev. D",
    volume = "59",
    pages = "025008",
    year = "1999"
}

@article{Bimonte:1998he,
    author = "Bimonte, G. and Iniguez, D. and Tarancon, A. and Ullod, C. L.",
    title = "{A Monte Carlo study of inverse symmetry breaking}",
    eprint = "hep-lat/9802022",
    archivePrefix = "arXiv",
    reportNumber = "DFTUZ-27-97, DSF-3-98",
    doi = "10.1103/PhysRevLett.81.750",
    journal = "Phys. Rev. Lett.",
    volume = "81",
    pages = "750--753",
    year = "1998"
}

@article{Jansen:1998rj,
    author = "Jansen, K. and Laine, M.",
    title = "{Inverse symmetry breaking with 4-D lattice simulations}",
    eprint = "hep-lat/9805024",
    archivePrefix = "arXiv",
    reportNumber = "CERN-TH-98-167",
    doi = "10.1016/S0370-2693(98)00775-8",
    journal = "Phys. Lett. B",
    volume = "435",
    pages = "166--174",
    year = "1998"
}

@article{Bimonte:1999tw,
    author = "Bimonte, G. and Iniguez, D. and Tarancon, A. and Ullod, C. L.",
    title = "{Inverse symmetry breaking on the lattice: An Accurate MC study}",
    eprint = "hep-lat/9903027",
    archivePrefix = "arXiv",
    reportNumber = "DFTUZ-98-27, DSF-7-99",
    doi = "10.1016/S0550-3213(99)00421-6",
    journal = "Nucl. Phys. B",
    volume = "559",
    pages = "103--122",
    year = "1999"
}

@article{Pinto:1999pg,
    author = "Pinto, Marcus B. and Ramos, Rudnei O.",
    title = "{A Nonperturbative study of inverse symmetry breaking at high temperatures}",
    eprint = "hep-ph/9912273",
    archivePrefix = "arXiv",
    reportNumber = "DFT-IF-UERJ-99-21",
    doi = "10.1103/PhysRevD.61.125016",
    journal = "Phys. Rev. D",
    volume = "61",
    pages = "125016",
    year = "2000"
}

@incollection{Dunne1999,
  author    = {Gerald V. Dunne},
  title     = "{Aspects of Chern--Simons Theory}",
  booktitle = {Topological Aspects of Low Dimensional Systems},
  pages     = {177--263},
  publisher = {Springer},
  year      = {1999},
  doi       = {10.1007/3-540-46637-1_3},
  eprint    = {hep-th/9902115},
  archivePrefix = {arXiv}
}

@article{Gross:1974jv,
    author = "Gross, David J. and Neveu, Andre",
    title = "{Dynamical Symmetry Breaking in Asymptotically Free Field Theories}",
    reportNumber = "COO-2220-19",
    doi = "10.1103/PhysRevD.10.3235",
    journal = "Phys. Rev. D",
    volume = "10",
    pages = "3235",
    year = "1974"
}

@article{Zinn-Justin:1991ksq,
    author = "Zinn-Justin, Jean",
    title = "{Four fermion interaction near four-dimensions}",
    reportNumber = "SACLAY-SPH-T-91-092",
    doi = "10.1016/0550-3213(91)90043-W",
    journal = "Nucl. Phys. B",
    volume = "367",
    pages = "105--122",
    year = "1991"
}

@article{Diab:2016spb,
    author = "Diab, Kenan and Fei, Lin and Giombi, Simone and Klebanov, Igor R. and Tarnopolsky, Grigory",
    title = "{On ${C}_{J}$ and ${C}_{T}$ in the Gross{\textendash}Neveu and O(N) models}",
    eprint = "1601.07198",
    archivePrefix = "arXiv",
    primaryClass = "hep-th",
    reportNumber = "PUPT-2496",
    doi = "10.1088/1751-8113/49/40/405402",
    journal = "J. Phys. A",
    volume = "49",
    number = "40",
    pages = "405402",
    year = "2016"
}

@article{Fei:2016sgs,
    author = "Fei, Lin and Giombi, Simone and Klebanov, Igor R. and Tarnopolsky, Grigory",
    title = "{Yukawa CFTs and Emergent Supersymmetry}",
    eprint = "1607.05316",
    archivePrefix = "arXiv",
    primaryClass = "hep-th",
    reportNumber = "PUPT-2504",
    doi = "10.1093/ptep/ptw120",
    journal = "PTEP",
    volume = "2016",
    number = "12",
    pages = "12C105",
    year = "2016"
}

@article{Goykhman:2020ffn,
    author = "Goykhman, Mikhail and Rosenhaus, Vladimir and Smolkin, Michael",
    title = "{The background field method and critical vector models}",
    eprint = "2009.13137",
    archivePrefix = "arXiv",
    primaryClass = "hep-th",
    doi = "10.1007/JHEP02(2021)074",
    journal = "{JHEP}",
    volume = "02",
    pages = "074",
    year = "2021"
}
